\documentclass[useAMS,usenatbib]{mnras}

\usepackage[english]{babel}
\usepackage{amsmath,amssymb}

\usepackage{graphicx}

\usepackage[normalem]{ulem}

\usepackage{verbatim}
\usepackage{color}
\usepackage{multirow}
\usepackage{mathtools}
\usepackage{epstopdf}

\usepackage{url}
\usepackage{xcolor}

\usepackage{gensymb}

\usepackage[usestackEOL]{stackengine}
\usepackage{xspace}

\usepackage{amsmath}
\usepackage{hyperref}

\def\nat{Nature}

\def\mnras{MNRAS}
\def\apj{ApJ}
\def\apjl{ApJL}
\def\apjs{ApJS}
\def\aap{A\&A}

\def\aj{AJ}

\def\physrep{Phys. Rep.}
\def\pasa{Publ. Astr. Soc. Australia}
\def\pasp{Publ. Astr. Soc. Pac.}

\def\pasj{Pub. Astron. Soc. Japan}

\def\be{\begin{equation}} 
\def\ee{\end{equation}} 
\def\ba{\begin{eqnarray}} 
\def\ea{\end{eqnarray}}

\def\cc{\,{\rm {cm^{-3}}}} 
\def\msun{{\Msun}}

\def\gsim{\lower.5ex\hbox{\gtsima}} 
\def\lsim{\lower.5ex\hbox{\ltsima}} \def\gtsima{$\; \buildrel > \over 
\sim \;$} \def\ltsima{$\; \buildrel < \over \sim \;$} \def\prosima{$\; 
\buildrel \propto \over \sim \;$} \def\gsim{\lower.5ex\hbox{\gtsima}} 
\def\lsim{\lower.5ex\hbox{\ltsima}} 
\def\simgt{\lower.5ex\hbox{\gtsima}} 
\def\simlt{\lower.5ex\hbox{\ltsima}} 
\def\simpr{\lower.5ex\hbox{\prosima}}   
  
 \def\gtsima{$\; \buildrel > \over \sim \;$} 
\def\ltsima{$\; \buildrel < \over \sim \;$} 
\def\gsim{\lower.5ex\hbox{\gtsima}} 
\def\lsim{\lower.5ex\hbox{\ltsima}} 
\def\simgt{\lower.5ex\hbox{\gtsima}} 
\def\simlt{\lower.5ex\hbox{\ltsima}} 
\def\simpr{\lower.5ex\hbox{\prosima}}

\def\msun{\,{\rm \Msun}}

\def\E3{{\cal E}_{\rm g}^{III}}

\def\r12{r_{1/2}} 
\def\x12{x_{1/2}} 
\def\v12{v_{1/2}}

\newcommand\code[1]{\textsc{\MakeLowercase{#1}}}

\def\nh2{n_{\rm H2}}
\def\fh2{f_{\rm H2}}

\def\arcsec{^{\prime\prime}}
\def\angstrom{\textrm{A\kern -1.3ex\raisebox{0.6ex}{$^\circ$}}}

\def\msun{{\rm M}_{\odot}}

\def\cc{{\rm cm}^{-3}}

\makeatletter
\def\@hex@@Hex#1%
 {\if a#1A\else \if b#1B\else \if c#1C\else \if d#1D\else
  \if e#1E\else \if f#1F\else #1\fi\fi\fi\fi\fi\fi \@hex@Hex}
\makeatother

\definecolor{apcolor}{HTML}{008000}
\definecolor{cbcolor}{HTML}{ff0f00}
\definecolor{afcolor}{HTML}{b3443c}
\definecolor{ddcolor}{HTML}{077a2f}
\definecolor{fdcolor}{HTML}{0e3466}
\definecolor{sgcolor}{HTML}{b3003b}
\definecolor{fvcolor}{HTML}{8900bf}

\defcitealias{Barai18}{B18}
\defcitealias{Gallerani10}{G10}

\newcommand{\lunit}{\mathrm{~erg ~s^{-1}}}

\newcommand{\cone}{\textit{AGNcone}\xspace}
\newcommand{\sphere}{\textit{AGNsphere}\xspace}

\newcommand{\noagn}{\textit{noAGN}\xspace}
\newcommand{\chandra}{\textit{Chandra}\xspace}
\newcommand{\nhunit}{\mathrm{ ~cm^{-2}}}

\newcommand{\athena}{\textit{Athena}\xspace}
\newcommand{\lynx}{\textit{Lynx}\xspace}
\newcommand{\axis}{\textit{AXIS}\xspace}
\date{}
\pagerange{\pageref{firstpage}--\pageref{lastpage}} \pubyear{2020}

\title[Feedback effect on the observable properties of $z>6$ AGN]{Feedback effect on the observable properties of $z>6$ AGN}

\author[]{
F. Vito,$^{1,2}$ 
F. Di Mascia,$^{1}$ 
S. Gallerani,$^{1}$ 
T. Zana,$^{1}$
A. Ferrara,$^{1}$
S. Carniani,$^{1}$
R. Gilli$^{2}$\\
$^{1}$Scuola Normale Superiore, Piazza dei Cavalieri 7, I-56126 Pisa, Italy\\
$^{2}$INAF -- Osservatorio di Astrofisica e Scienza dello Spazio di Bologna, Via Gobetti 93/3, I-40129 Bologna, Italy\\
}

\begin{document}

\maketitle
\label{firstpage}
\begin{abstract}
 Active galactic nuclei (AGN) feedback has a major impact onto the supermassive black-hole (SMBH) growth, the properties of the host galaxies, and their cosmic evolution. We investigate the effects of different kinetic feedback prescriptions on the observable properties of AGN and their host galaxies at $z>6$ in a suite of zoom-in cosmological simulations. We find that kinetic feedback decreases the column density of the interstellar medium (ISM) in the host galaxy by up to a factor of $\approx10$, especially when the SMBHs reach high accretion rates ($\approx10-30\,\mathrm{M_\odot\,yr^{-1}}$). In particular, kinetic feedback is required to extend the ISM size to $>1$ kpc and match the observed sizes of the gas reservoirs in $z>6$ AGN host galaxies. Moreover, it produces unobscured lines of sight along which the AGN can be detected in the rest-frame UV band with magnitudes consistent with observed values of $z>6$ AGN. The assumed geometry of the outflow plays an important role in shaping the observed properties of high-redshift AGN. We find that a biconical geometry is favored over a spherical one to reproduce the observed properties, but it overestimates the number of multiple AGN systems detectable in X-ray observations. This result suggests that simplistic BH seeding recipes widely employed in cosmological simulations produce too many X-ray detectable multiple AGN at $z=6-7$, thus soliciting the adoption of more physically motivated seeding prescriptions.
\end{abstract}


\section{Introduction}

The existence of super-massive black holes (SMBHs) with masses larger than billion solar masses at $z\gtrsim6$ \citep[e.g.,][]{Banados18a,Matsuoka18a, Wang21b}, when the Universe was $<1$ Gyr old, challenges our current understanding of SMBH and galaxy formation and evolution, and is thus one of the most pressing open issues in modern astrophysics \citep[e.g.,][]{Woods19}.
Their distance and faintness make observations of these objects difficult and strongly biased towards the most luminous and massive accreting SMBHs. A complementary approach is to use numerical simulations as tools to study the largely unknown phases of SMBH growth in the early Universe \citep[e.g.,][]{Tanaka09,Sijacki09, Habouzit16, Habouzit19}. 
However, observed properties of high-redshift accreting SMBHs, or active galactic nuclei (AGN), and predictions of numerical simulations have been compared only seldom \citep[e.g.,][Zana et al. accepted]{Ni20, Habouzit21, DiMascia21a}.

An important ingredient entering in numerical simulations focused on the early growth of SMBHs is the effect of AGN feedback \citep[e.g.][]{Costa14, Costa20, Barai18, Habouzit19, Valentini21}, as it is often considered to have a major role in shaping the evolution of AGN and galaxies along the whole cosmic history \citep[e.g.,][]{Fiore17}. In particular, optically-selected luminous quasi-stellar objects (QSOs) in the early Universe often present evidence for the launching of fast and massive multi-phase outflows (e.g., \citealt{Maiolino12, Cicone15,Bischetti19, Carniani19, Schindler20, Izumi21}; but see also, e.g., \citealt{Decarli18, Novak20, Meyer22}), which are expected to affect the observable properties of the QSOs themselves and their host galaxies, such as X-ray oscuration, UV extinction, and gas content \citep[e.g., ][]{Brusa15b, Ni20}.

Outflows observed in QSOs are though to originate from fast nuclear winds, which, in turn, may be accelerated by several physical mechanisms, including radiation pressure, due to UV photons produced in the accretion disc, on dust grains or on partially ionized gas mediated by UV transitions, and magnetic effects \citep[e.g.][]{Proga00, Murray05, Fabian08, Yuan15, RicciC17}. The physical scales involved in these processes are those of the accretion disk \citep[e.g., ][]{Giustini19}. Since such scales cannot be resolved by large-scale cosmological simulations, different authors have modeled AGN feedback 
using several different recipes (e.g., \citealt{Barai18, Costa20, Ni20}).
Moreover, the effect of the outflow on the surrounding material can potentially depend on its geometry \citep[e.g.,][]{Zubovas16}. 
Since the exact acceleration physics, and thus launching direction, of nuclear winds is not well understood, numerical simulations typically assume either spherical \citep[e.g., ][]{Feng16} or bi-conical \citep[e.g., ][]{Sala21} outflow geometry as study cases.

Beside the properties of the individual galaxies hosting accreting SMBHs, numerical simulations provide also information on the environment of high-redshift luminous AGN. While these objects are expected to reside in the peaks of the dark matter halo distribution, which are generally characterized by large overdensities of galaxies (e.g., \citealt{Costa14, Wise19}), although with some scatter (e.g., \citealt{Habouzit19}), observations struggle to provide us with a clear view of typical high-redshift QSO environment. In fact, $z>6$ QSOs have been reported to reside in a variety of environments, including underdense, normal, and overdense regions (e.g. \citealt{Ota18, Mazzucchelli19,Overzier21}). The first spectroscopically confirmed galaxy overdensity around a $z>6$ QSOs was presented recently by \cite{Mignoli20}, followed by a tentative confirmation of another structure by \cite{Overzier21}. 

A significant fraction ($\approx40\%$) of $z\gtrsim6$ QSOs has ALMA-detected dusty companion galaxies at distances of a few kpc \citep[e.g.][]{Willott17, Decarli18, Neeleman19, Venemans20}. These satellite galaxies might host heavily reddened and buried AGN \citep[e.g., ][]{DiMascia21a}, although currently there is no strong observational evidence for the presence of accreting SMBHs in their centres \citep[e.g., ][]{Connor19,Connor20, Vito19a, Vito21}. 
Such objects would be typically brighter than inactive galaxies, expecially in the X-ray band. Therefore, their predicted number in numerical simulations can be tested against observational results to infer how well simulations approximate reality.

In this paper, we present a study of the effect of AGN kinetic feedback on the observable properties of $z>6$ AGN in cosmological simulations. In particular, we analyse a set of numerical simulations presented by \citet[][hereafter, \citetalias{Barai18}]{Barai18} with different kinetic feedback prescriptions, focusing on the most massive SMBH at $z=6$ and its surrounding environment. We extract multiwavelength observables such as column density and radial extent of the gas distributed in the host galaxies, UV and X-ray AGN fluxes, and number of satellite AGN detectable over small (i.e., a few kpc) distances from the central SMBH. We compare these properties with results from multiwavelength observations.

The paper is structured as follows.
In \S~\ref{Method} we describe the numerical setup of the simulations, the AGN selection, and the method used to measure the gas column density and distribution. In \S~\ref{NH_distro} we discuss the redshift evolution of the column densities for the considered AGN. In \S~\ref{comparison_obs} we present the observable properties of the simulated AGN and their host galaxies, and we compare them with empirical findings. In \S~\ref{environment} we investigate the presence of multiple AGN systems over scales of a few kpc, and we compare their detectability rates in the X-ray band with results from observations of high-redshift AGN. Finally, in \S~\ref{discussion} we discuss and interpret the results, and in \S~\ref{conclusions} we provide a summary.
All quoted distances are physical unless otherwise noticed. 
We adopt a flat $\Lambda$CDM cosmology with $H_0=67.7\,\mathrm{km\,s^{-1}}$ and $\Omega_m=0.307$ \citep{Planck16}.

\section{Method}\label{Method}
\subsection{Numerical model} \label{Numerical_methods}

We consider the simulation runs \cone and \sphere by \citetalias{Barai18}, which include kinetic feedback.  We provide here a summary of the numerical setup and refer to the original works for an in-depth discussion.

{\citetalias{Barai18}} used a modified version of the Smooth Particle Hydrodynamics (SPH) N-body code \code{GADGET-3} \citep{Springel05} to follow the evolution of a comoving volume of ($500$ Mpc)$^3$, starting from cosmological initial condition generated with \code{music} \citep{hahn11} at $z=100$, and zooming-in on the most massive (i.e., $4\times10^{12}\,\mathrm{M_\odot}$) dark matter (DM) halo, corresponding to a $\approx3\sigma$ overdensity \citep[e.g.,][]{Barkana01}, inside the box down to $z=6$. Therefore, the final zoomed-in simulations focus by construction on a highly biased cubic region, with a volume of (5.21 Mpc)$^3$. The highest level of the simulation has a mass resolution of $m_{\rm DM} = 7.54 \times 10^6$ $\msun$ and $m_{\rm gas} = 1.41 \times 10^6$ $\msun$ for DM and gas particles, respectively. The softening length for gravitational forces for these high-resolution DM and gas particles is $R_{\mathrm{soft}} = 1 h^{-1}$ ckpc.

The code accounts for gas heating and cooling (including metal-line cooling) depending on the gas metal content, based on eleven element species (H, He, C, Ca, O, N, Ne, Mg, S, Si, Fe) that are tracked in the simulation \citep{Tornatore07}. Star formation in the inter-stellar medium (ISM) is implemented following the multiphase effective subresolution model by \citet{Springel03}, adopting a density threshold for star formation of $n_{SF} = 0.13 \ \cc$.
The simulations include stellar winds, supernovae feedback, and metal enrichment, and assume a \citet{Chabrier03} initial mass function in the mass range $0.1-100$ $\msun$ \citep{Tornatore07,barai13,biffi16}. 

When a DM halo that is not already hosting a black hole (BH) reaches a total mass of $M_{\rm h} \geq 10^9$ $\msun$, a $M_{\rm BH} = 10^5$ $\msun$ BH is seeded at its centre. BHs are treated as collisionless sink particles and are allowed to grow by accretion of the surrounding gas or by mergers with other BHs. Gas accretion onto BHs is modelled via the classical Bondi-Hoyle-Littleton accretion rate $\dot{M}_{\rm Bondi}$ \citep{Hoyle39, Bondi44, Bondi52}, capped at the Eddington rate $\dot{M}_{\rm Edd}$:
\begin{equation}
    \dot{M}_{BH} = {\rm min} (\dot{M}_{\rm Bondi}, \dot{M}_{\rm Edd}).
\end{equation}
Accreting BH radiate away a fraction $\epsilon_{\rm r}$ of the accreted rest-mass energy, with a bolometric luminosity
\begin{equation}\label{eq:luminosity_bh}
    L_{\rm bol} = \epsilon_{\rm r} \dot{M}_{\rm BH} c^2,
\end{equation}
where $c$ is the speed of light. \citetalias{Barai18} fixed the radiative efficiency to $\epsilon_{\rm r} = 0.1$, a fiducial value for radiatively efficient, geometrically thin, optically thick accretion disks around a Schwarzschild BH \citep{Shakura73}. 

A fraction $\epsilon_{\rm f} = 0.05$ of the total output energy is distributed to the surrounding gas in a kinetic form\footnote{We refer to \citetalias{Barai18} for details about the choice of the value for $\epsilon_{\rm f}$ and the numerical implementation of the kinetic feedback.}.  In \cone the kinetic energy is distributed along two cones with a half-opening angle of $45\degree$. The direction of the cone axis is chosen randomly for each BH at the seeding time, and is kept fixed throughout the simulation \citep{Barai18}, similarly to what is done in \cite{Zubovas16}. Instead, the AGN feedback in \sphere pushes away the gas particles along random directions, thus mimicking a spherical geometry.

\begin{figure*}
    \centering
    \includegraphics[width=1\textwidth ]{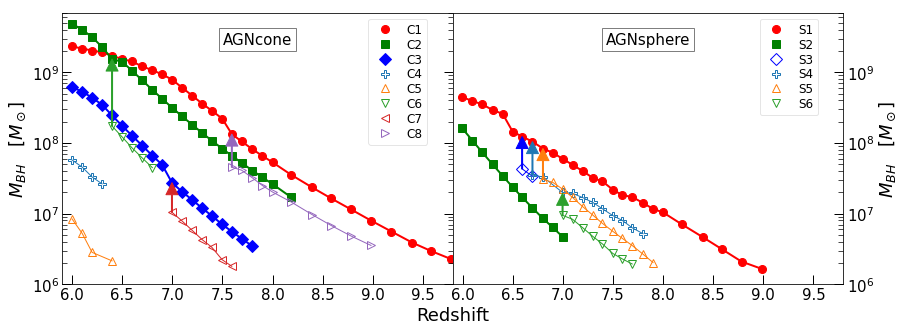}
    
    \caption{BH masses as a function of redshift for the  \cone (left) and \sphere (right) runs. Only SMBHs accreting at $\dot{M}>0.02\,\mathrm{M_\odot}$ are considered. The arrows mark the mergers between BHs. AGN considered in \S~\ref{NH_distro} and \S~\ref{comparison_obs} (i.e., those that reach $z=6$ with $M_{BH}>10^8\,\mathrm{M_{\odot}}$) are plotted as filled symbols.}
    %In the \cone run, the three most massive BHs at $z=6.3$ are color-coded as in \citet{DiMascia21a}. \FD{I would remove this sentence.}} 
    \label{fig:masses}
\end{figure*}

\begin{figure*}
    \centering
    \includegraphics[width=1\textwidth ]{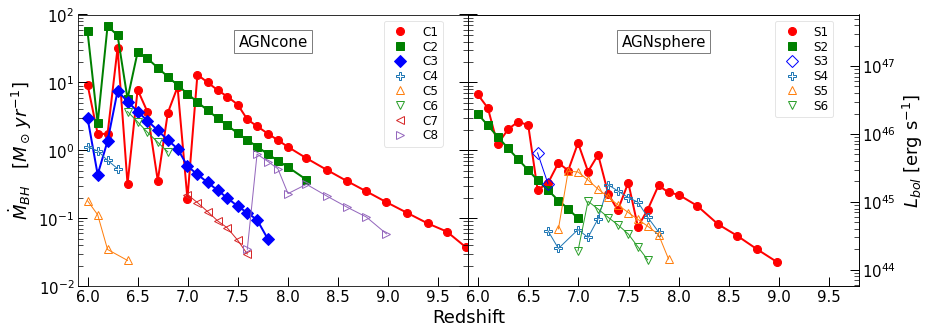}
    
    \caption{Mass accretion rate as a function of redshift for the  \cone (left) and \sphere (right). The corresponding bolometric luminosity (Eq.~\ref{eq:luminosity_bh}) is reported in the right axis.  }
    \label{fig:mdot}
\end{figure*}

\subsection{AGN selection}\label{selection}

We analyse the simulation snapshots in steps of $\Delta z =0.2$ from $z=10$ to $z=8$ and $\Delta z =0.1$ from $z=8$ to $z=6$. In particular, we follow the most massive SMBH at $z=6$ in each simulation set, and consider a box with side size of 60 kpc centred on it. We refer to all of the SMBHs in the box accreting at $\dot{M_{BH}}>0.02\,\mathrm{M_\odot\,yr^{-1}}$ (i.e., $L_{bol}\approx10^{44}\lunit$) as AGN. Fig.~\ref{fig:masses} presents the BH mass evolution of AGN in the two simulations. Each AGN is labelled with the initial letter of the run (C for \cone, S for \sphere). %The first AGN in each run (e.g., C1 in \cone) is hosted in the central and most massive galaxy of the structure. 

\cone forms two very massive ($>10^9\,M_\odot$) BHs at $z<7$, while only less massive BHs are formed in the \sphere run. This behaviour is linked to the implementation of the feedback: \cone allows the gas to accrete continuously along the equatorial directions, while the lack of a preferential direction along which the outflow is launched in \sphere does not allow for a steady and efficient accretion onto the SMBH. This effect can be appreciated in Fig.~\ref{fig:mdot}: the accretion rate of \cone is generally higher than that of \sphere, at least up to $\dot{M}\approx1-30\,\mathrm{M_\odot\,yr^{-1}}$. At higher accretion rates, which are reached by the most accreting BHs at $z<7$, AGN feedback prevents further increase of the accretion rate. 

Hereafter, we focus our analysis on the AGN that reach $z=6$ with $M_{BH}>10^8\,\mathrm{M_\odot}$ and $L_{bol}>10^{46}\,\mathrm{erg\,s^{-1};}$ (see filled symbols in Fig.~\ref{fig:masses} and Fig.~\ref{fig:mdot}), which we refer to as ``bright AGN" (i.e., C1, C2, and C3 in \cone; S1 and S2 in \sphere).
These  BH mass and luminosity values are typical of known $z>6$ QSOs \citep[e.g., ][]{Yang21}, allowing us to compare the physical properties of simulated and observed AGN in a consistent way. We note that, since the simulations focus on a single cosmic region at high redshift, the derived expectations on the AGN observable properties might be affected by cosmic variance.

\subsection{Gas column density and radial distribution}\label{NH}
Here we describe the method that we use to derive the distribution of hydrogen, helium, and metal column densities in the ISM for galaxies hosting AGN in the considered simulations. We make use of the hydrogen column density in the remaining of the paper to derive the observational properties predicted by the two considered simulations.

\begin{figure*}
    \centering
    \includegraphics[width=1\textwidth ]{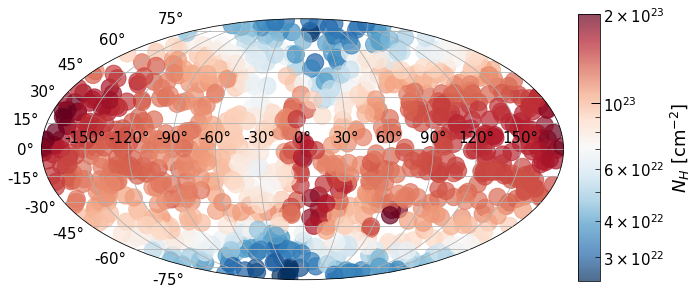}
    \includegraphics[width=1\textwidth ]{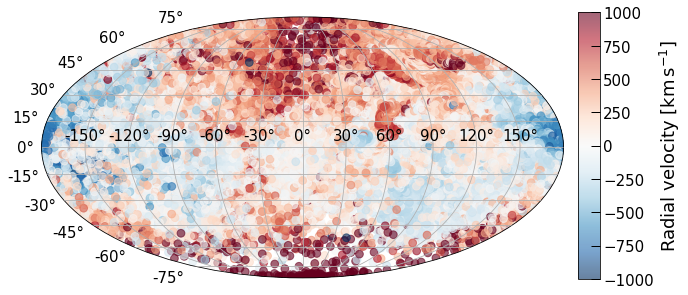}  
    \caption{\textit{Upper panel}: Mollweide projection of the column density along 1000 random lines of sights centred on the QSO C1 at $z=7.1$. \textit{Lower panel}: Mollweide projection of the radial velocities of all particles within 10 kpc from C1 at $z=7.1$. The different sampling of the maps is intended to show the homogeneity of the 1000 LOSs used to compute $N_H$ in the upper panel, and the velocity of the individual gas particles in the lower panel. The map is aligned with the outflow cone direction. Regions where the particles have high positive velocities correspond to the two cones along which the kinetic energy is distributed by the AGN feedback in the \cone simulation. Such cones are characterised by the lowest values of column densities.}
    \label{fig:Mollweide}
\end{figure*}

We estimate the distribution of the column densities for the bright AGN in the simulations by launching 1000 randomly selected lines of sight (LOSs) toward each AGN from a distance $d= 30$ kpc. Each LOS is considered as the axis of a cylinder with basis radius of $R_{\mathrm{soft}}$. We note that the resolutions of the simulations do not allow us to probe structures on smaller scales, as, for instance, the existence of a dusty torus on pc scales. Then, each cylinder is divided along its length into bins of $l_{\mathrm{bin}}=0.25$ kpc width, for a total of $\frac{d}{l_\mathrm{bin}}=120$ radial bins. We compute the density of each chemical element in a bin of the cylinder from the mass carried by each particle included in that bin.
With this approach, we also obtain the radial distribution of the gas density.
%Following \cite{Ni20}, we assumed that the gas in particles with star-formation rate $SFR>0$ is neutral.
Finally, we integrate along the cylinder to compute the total column density of hydrogen ($N_H$) and of the other elements. The resulting total $N_H$ is not sensitive to reasonably different values of $l_{\mathrm{bin}}$ (i.e., from 0.25 kpc to 1 kpc). Therefore, we used $l_{\mathrm{bin}}=0.25$ as this value allows us to sample well the radial distribution of the gas (see \S~\ref{radial_distro}).
Fig.~\ref{fig:Mollweide} (upper panel) presents an example of the derived column-density map centred on the QSO C1 in \cone. Each circle represents one of the 1000 random LOSs, which sample homogeneously the entire solid angle as seen from C1.

To assess the effect of feedback on the column density (\S~\ref{NH_distro}), we also consider an additional simulation run presented in \cite{Barai18}, that is identical in terms of initial conditions and physical prescriptions to the \cone and \sphere, except that BHs are not seeded. The only type of feedback in this run, which we refer to as \noagn, is due to supernovae explosions (see \citealt{Barai18} for detailed discussion). 
 
 We associate each AGN in a simulation to the corresponding galaxy in the \noagn runs following a method similar to that described in Zana et al. (accepted): first, we identify the DM halo hosting the AGN as the one having its centre of mass closest to the position of the SMBH. Then, we identify the corresponding halo in the \noagn run by cross-matching the DM particle IDs in the two runs, and selecting the halo in \noagn which shares the largest fraction of particles with the initial AGN halo, further imposing that the mass difference must be within a factor of 10-50.\%\footnote{The exact threshold is adjusted at each time step in order to find at least one halo counterpart.} Finally, we repeat the procedure described above on the selected halo in \noagn, and derive the column density distribution in absence of AGN feedback. At $z>8$, the redshifts at which the \noagn snapshots are taken are significantly different from those of the runs including AGN, making the DM-halo match procedure highly uncertain. Thus, we limit the identification of the AGN-hosting galaxies counterparts in the \noagn run to $z<8$.
 
 \begin{figure*}
    \centering
    \includegraphics[width=1\textwidth ]{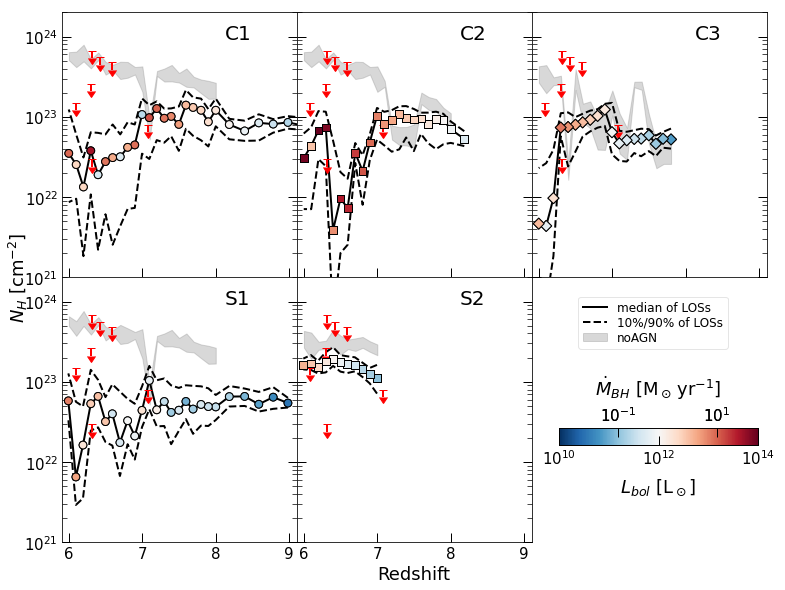}
    \caption{Evolution of column density for bright AGN in the \cone (C1, C2, C3) and \sphere (S1, S2) simulations. We show the median value (solid line, color coded according to the AGN bolometric luminosity and accretion rate), and the 10\% and 90\% percentiles (dashed lines) computed by launching 1000 lines of sight. The gray stripes enclose the 10\% to 90\% percentiles of the column densities of matched galaxies in the same simulation sets where, however, BHs have not been seeded (i.e., the \noagn case). To compare with observational results (\S~\ref{Xray_obsc}), the red arrows mark the 3$\sigma$ upper limits derived for X-ray detected QSOs with $>10$ counts from \citet{Nanni18} and \citet{Connor19}.}
    \label{fig:NH_z_all}
\end{figure*}

   \begin{figure*}
    \centering
    \includegraphics[width=1\textwidth ]{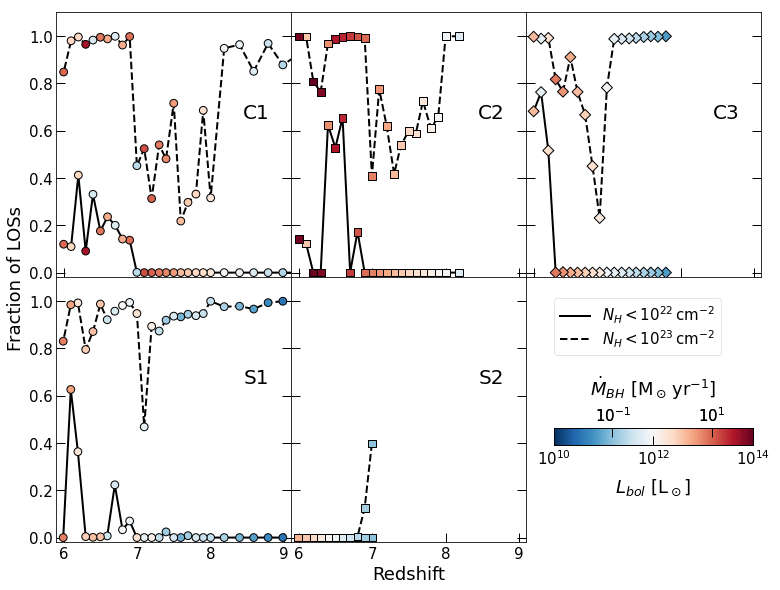}
    \caption{Fraction of lines of sight obscured by column densities $<10^{22}\,\mathrm{cm^{-1}}$ (solid lines) and $<10^{23}\,\mathrm{cm^{-1}}$ (dashed lines) as a function of redshift for the bright AGN in the \cone (C1, C2, C3) and \sphere (S1, S2) simulations. The symbols are color coded according to the AGN bolometric luminosity and accretion rate.}
    \label{fig:fLOS_all}
\end{figure*}
 
 \section{Column density evolution}\label{NH_distro}
 Fig.~\ref{fig:NH_z_all} presents the evolution of the column density for bright AGN in the \cone and \sphere simulations.
Considering the \cone simulation, the AGN column densities are similar to, or slightly lower than, those derived for the corresponding galaxies in the \noagn run until the AGN accretion rate reaches $\dot{M}\approx10-30\,\mathrm{M_\odot\,yr^{-1}}$. This happens at $z\approx7$ for C1 and C2, and $z\approx6.3$ for C3 (see Fig.~\ref{fig:mdot}). At later times, the AGN column density drops significantly by up to $\approx1$ dex and the accretion rate starts to oscillate. The 10\% and 90\% percentiles span up to one order of magnitude, especially at $z=6-7$, when the accretion rates reach the maximum values, producing the most powerful conical outflows.

Instead, the column densities of the corresponding galaxies in \noagn  (grey stripes in Fig.~\ref{fig:NH_z_all}) keep on increasing relatively smoothly. This finding confirms the AGN $N_H$ drop and the presence of unobscured LOSs to the effect of the conical kinetic feedback. At low accretion rates the produced outflow cannot stop the infall of material, but once the accretion rate reaches high enough values, the energy carried by the outflow impacts a significant part of the gas in the halo, hindering further infalling, especially along the conical outflow directions. As a result, the $N_H$ decreases, as well as the AGN accretion rate, until more material is allowed to accrete, producing a new burst of powerful feedback. Such a cyclic activity explains the decreasing median $N_H$, the wider $N_H$ distribution, and the oscillating $\dot{M}$ behaviour at later cosmic times. 
This result is in qualitative agreement with the self-regulation scenario  discussed by, e.g.,  \cite{Sijacki09, Dubois13, Costa14, Feng14,Richardson16, Trebitsch19}, according to which the AGN feedback controls the growth of the black hole and limits the duration of high accretion episodes by emptying the host galaxy gas reservoir, provided that the accretion rate is sufficiently high.

However, we note that the physical interpretation of our results is complicated by the effect that one AGN may have on other AGN-hosting galaxies passing through its feedback cone. In fact, C1, C2, and C3 in \cone at $z<7$ are always closer than 30 kpc, and reach minimum distances as small as 4 kpc. At these distances, powerful outflows launched from one AGN may affect nearby galaxies (e.g., Zana et al. accepted).

As an example of the feedback effect on the column density, in Fig.~\ref{fig:Mollweide}, we compare the $N_H$ map centred on C1 with the radial velocity map of all particle within 10 kpc from C1. The maps correspond to $z=7.1$, when C1 reaches a local maximum in accretion rate before the strong AGN feedback starts to impact significantly the $N_H$ (Fig.~\ref{fig:NH_z_all}) and $\dot{M}$ starts to oscillate. Comparing the column density map (upper panel) with the map of the radial velocity of individual particles (lower panel), we notice that the two conical outflows, identified as regions with positive radial velocities, correspond to LOSs with low column densities. Such LOSs are those along which high-redshift AGN are more easily to be detected in the rest-frame UV band, as we investigate in details in \S~\ref{UV}.
Fig.~\ref{fig:fLOS_all} presents the fraction of LOSs along which $N_H<10^{22}\,\nhunit$ (solid lines) and $N_H<10^{23}\,\nhunit$ (dashed lines) for each bright AGN. 
 Hereafter, we use the widely used threshold $N_H=10^{22}\,\nhunit$ to separate obscured and unobscured AGN.\footnote{ However, we note that we consider the dust extinction as a more relevant quantity when we study the AGN rest-frame UV emission in \S~\ref{UV}.} For instance, \cite{Merloni14} found that such a value returns the best agreement between samples of obscured AGN as defined in optical (e.g., narrow emission-line AGN) and X-ray bands.  
From Fig.~\ref{fig:fLOS_all} we infer that only at $z\lesssim7$ a fraction of the LOSs would appear as unobscured. In particular, at $z\lesssim7$ C1 presents unobscured LOSs over $10-40\%$ of the solid angle, while this fraction is much more variable with redshift (i.e., $0-80\%$) for C2 and C3.

The most massive BH in the \sphere simulation, S1, follows a somewhat similar $N_H$ evolution to that of the AGN in \cone: a roughly constant median $N_H$ value up to $z\approx7$ followed by a slightly decreasing and wider $N_H$ distribution (Fig.~\ref{fig:NH_z_all}), and the appearance of unobscured LOSs (Fig.~\ref{fig:fLOS_all}) at later cosmic times. However, some differences exist: first, the AGN $N_H$ is always significant lower than that of the corresponding galaxy in the \noagn run (grey stripe), even at $z>7$. Secondly, the column density drop at $z<7$ is not as strong as in the \cone case. Finally, at $z>7$ the accretion rate of S1 is not as smooth as in the \cone case, and keeps on increasing even at $z<7$. 
These differences may be due to the prescripted geometry of the kinetic feedback in the \sphere case, in which gas particles are accelerated in a random direction during every accretion event. Therefore, in contrast with the \cone case, there is no preferential direction (i.e., the equatorial plane of the conical outflow) along which material can keep on accreting undisturbed for long periods of time at $z>7$. In particular, the accretion rate of S1 never exceeds $\approx10\,\mathrm{M_\odot\,yr^{-1}}$, which is the approximate threshold after which the AGN kinetic feedback affects more evidently the $N_H$ distribution and the accretion rate of AGN in the \cone run. 

The column density evolution of S2, instead, does not appear to be strongly influenced by the AGN feedback. Although the median $N_H$ is slightly lower than the values found in the \noagn case, it remains constant with time, and does not drop even at $z<6.5$, when S2 reaches similar accretion rate to S1. As a result, S2 would never appear as an unobscured AGN. We note that the typical column density of S2 is a factor $\approx3$ higher than that of S1 at any redshift, and its accretion rate rises smoothly from $z=7$ to $z=6$. These properties suggest that higher accretion rates than the values reached by S2 are required in order to launch outflows powerful enough to sweep away the gas in the case of large column densities (e.g., \citealt{Trebitsch19}), even when kinetic energy is distributed along random directions by the AGN feedback. 
%The evolution of $N_H$ for S2 is similar to that found by  \cite{Ni20} for three out of four luminous AGN at $z=7-10$ in the BlueTides simulation, although they are characterized by larger values of column densities (i.e., up to $N_H=10^{24}\,\nhunit$ at $z=7-8$).
 
The median values of $N_H$ we derive from the \cite{Barai18} simulations are consistent with typical values found by \cite{Lupi22}. However, the resolution of that work is $\times 85$ higher than our simulations, and allows the authors to sample compact regions of dense gas with $N_H\gtrsim10^{24}\,\mathrm{cm^{-2}}$, especially at $z>8$, when AGN feedback has not yet affected significantly the ISM distribution and density in the host galaxies. One of the main methodological differences with that work is that we compare the ISM densities in the same galaxies in which SMBHs are actively accreting or are not seeded at all. Thus, we probe directly the effect of AGN feedback on the ISM in the host galaxy. 

\section{Comparison with observations}\label{comparison_obs}
 In this section, we compare the observable properties derived from the $N_H$ distributions of the AGN predicted by the simulations (\S~\ref{NH_distro}) with observational results. In particular we focus
 on the comparison with constraints from X-ray observations  (\S~\ref{Xray_obsc}),
 the radial distribution of the gas reservoirs (\S~\ref{radial_distro}), and the observed UV magnitudes (\S~\ref{UV}).

\subsection{X-ray obscuration}\label{Xray_obsc}
X-ray observations are routinely used to constrain the column density of obscuring material along the LOSs of AGN.
Low and moderate values of column densities (\mbox{$N_H\lesssim10^{22}\,\nhunit$}) can absorb soft X-ray photons (rest-frame energies $\lesssim2$ keV), whereas larger column densities are required to absorb a high fraction of more energetic photons. 
However, X-ray observations of high-redshift QSOs \citep[e.g.][]{Vito19b,Wang21a} sample rest-frame energies $E>3$ keV, and are thus sensitive only to high column densities ($N_H\gtrsim 3\times10^{23}\,\nhunit$), at least at the sentivities of currently available facilities. Moreover, all of the known $z>6$ QSOs have been selected based on their unobscured rest-frame UV emission (i.e., they are optically classified as type 1 QSOs), and thus are not expected to be heavily obscured in the X-ray band. For these reasons, existing X-ray observations of bright $z>6$ QSOs provide us with only loose upper limits of $N_H$. The downward-pointing red arrows in Fig.~\ref{fig:NH_z_all} are the observed upper limits on $N_H$ derived for a sample of $z>6$ QSOs by \citealt{Nanni17} and \citealt{Connor19}, with typical luminosities $L_{bol}=10^{46}-10^{47}\,\mathrm{erg\,s^{-1}}$. The column densities derived for bright AGN in all of the considered simulations are lower than, or consistent with, such loose upper limits. Although the $N_H$ values found for the \noagn case are typically higher, they are still consistent with some measured upper limits. Therefore, the constraints on $N_H$ obtained from X-ray observations of high-redshift QSOs only marginally favour the presence of kinetic feedback. 

We note that constraining AGN obscuration using X-ray observations requires an assumption on gas metallicity, as X-ray photons are mainly absorbed by metal atoms. Typically, solar metallicity is assumed, whereas the ISM metallicity of the host galaxies of the AGN in the \citetalias{Barai18} simulations is sub-solar (e.g., by factors of $\approx2-3$ at $z=6$; e.g., Zana et al. in prep.). This consideration reinforces the overall consistency between the $N_H$ values constrained from X-ray observations and found in the simulations, as significantly larger column densities would be required in the case of sub-solar metallicities to produce X-ray obscuration in excess to that observed in real QSOs. In \S~\ref{environment} we discuss the X-ray detectability of the QSOs in the simulations.

\begin{figure*}
    \centering
    \includegraphics[width=1\textwidth ]{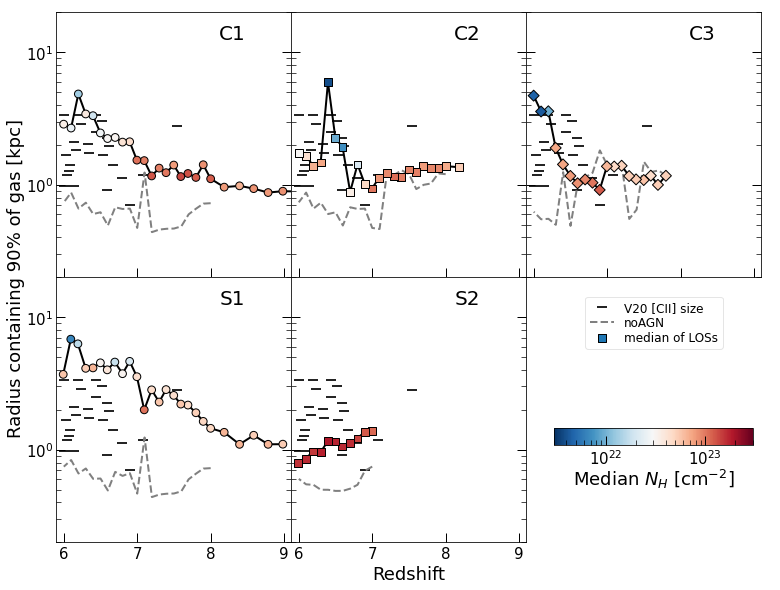}
    \caption{Median radius ($R_{90}$, computed over all of the lines of sight) containing the $90\%$ of the total gas as a function of redshift. The color code indicates the median total $N_H$, averaged over all of the lines of sight. The dashed grey lines mark the same quantity computed for matched galaxies in the same simulation sets where, however, BHs have not been seeded (i.e., the \noagn case).  The black ticks mark $R_{90}$ for 25 $z>6$ QSOs, as estimated from the [C II] emission beam-deconvolved sizes presented by \citet{Venemans20}.}
    \label{fig:NH_radius}
\end{figure*}

\subsection{Gas radial distribution}\label{radial_distro}
We investigate the effect of kinetic feedback on the observable sizes of the gas reservoirs in high-redshift QSOs. From the radial distribution of $N_H$ derived
for each LOS in \S~\ref{Method}, we computed the radius from the centre of the galaxy which includes $90\%$ of the gas contributing to the total $N_H$. Then, for each galaxy, we computed the median value considering all of the 1000 LOSs, and define it as $R_{90}$. We use such a quantity to quantify the size of the gas reservoir in a galaxy.

Fig.~\ref{fig:NH_radius} presents $R_{90}$ as a function of redshift for every bright AGN in the \cone and \sphere simulations, as well as for the matched galaxies in the \noagn runs. All of the bright AGN in the \cone simulation (C1, C2, C3) have a similar evolution of $R_{90}$: their gas reservoir sizes are constant ($\approx1$ kpc) at $z\gtrsim7$. At lower redshift, where $N_H$ decreases due to strong effect of the kinetic feedback, which is proportional to $\dot{M}$ and $L_{bol}$ (see the color-code of the circles in Fig.~\ref{fig:NH_z_all} and Fig.~\ref{fig:NH_radius}), $R_{90}$ increases up to several kpc. This behaviour is expected considering that the AGN feedback applies a mechanical push to the surrounding gas particles. In fact, the size of the gas reservoir in the \noagn run, where the AGN feedback lacks (grey dashed lines in  Fig.~\ref{fig:NH_radius}), remains constant or tends to even decrease at later cosmic times. 

The evolution of $R_{90}$ for S1 in the \sphere simulation is similar to that of the AGN in the \cone simulation. However, the increase of $R_{90}$ is stronger and begins at earlier cosmic times. We recall that the accretion rate of S1 is typically lower than that of the AGN in \cone (see Fig.~\ref{fig:mdot}), and therefore the stronger evolution of $R_{90}$ is not due to intrinsically stronger outflows launched by the AGN, but, as discussed in \S~\ref{NH_distro}, to the different geometry of the outflow: being launched along random directions at every accretion event, the outflow is more likely to transmit the kinetic energy to the gas particles in the galaxy even at low or moderate accretion rates. Instead, S2 does not follow the same evolution as S1. On the contrary, $R_{90}$ decreases to sub kpc values approaching $z=6$. As discussed in \S~\ref{NH_distro} we ascribe this behaviour to the relatively low accretion rate, which does not produce feedback strong enough to efficiently affect the gas distribution in the host galaxy.

We compare our findings with the observed extent of the [C II] emission of 25 $z>6$ QSOs presented by \cite{Venemans20}, assuming that the [C II] emission line is a good tracer of the spatial extent of the total gas reservoir \citep[e.g.,][]{Zanella18, Sommovigo21}. We used the major axis of the deconvolved [C II] emission size (Tab. 3 of \citealt{Venemans20}), which represents the FWHM of the emitting source, and converted it into the radius that includes 90\% of the [C II] light, assuming a Gaussian distribution.\footnote{We note that the conclusions hold if we use an exponential profile (e.g., \citealt{Fujimoto20}) and convert the FWHM values reported by \cite{Venemans20} into exponential scale lengths. In this case, we obtain larger radii than in the Gaussian case by a factor of $\approx1.75$.} The resulting values are reported in 
Fig.~\ref{fig:NH_radius}) as black ticks at the redshift of each QSO.

The AGN  in the \cone simulation have $R_{90}$ consistent with the observed values, while the median gas size of S1 is larger at nearly every redshift. S2 have a size consistent with the most compact QSOs in the \cite{Venemans20} sample. However, this comparison is not fair: the ISM in S2 produces very large column densities at all redshifts and all LOSs (Fig.~\ref{fig:NH_z_all}), and thus large expected values of dust extinction. All of the QSOs studied in \cite{Venemans20} are instead rest-frame UV selected objects: we lack observational information about the extent of the gas reservoirs of buried high-redshift QSOs, as is S2.  In all cases, the median gas size of the \noagn control galaxies are smaller than the observed values for QSOs, suggesting that kinetic feedback is required to produce the gas extents observed in real QSOs.

\begin{figure*}
    \centering
    \includegraphics[width=1\textwidth ]{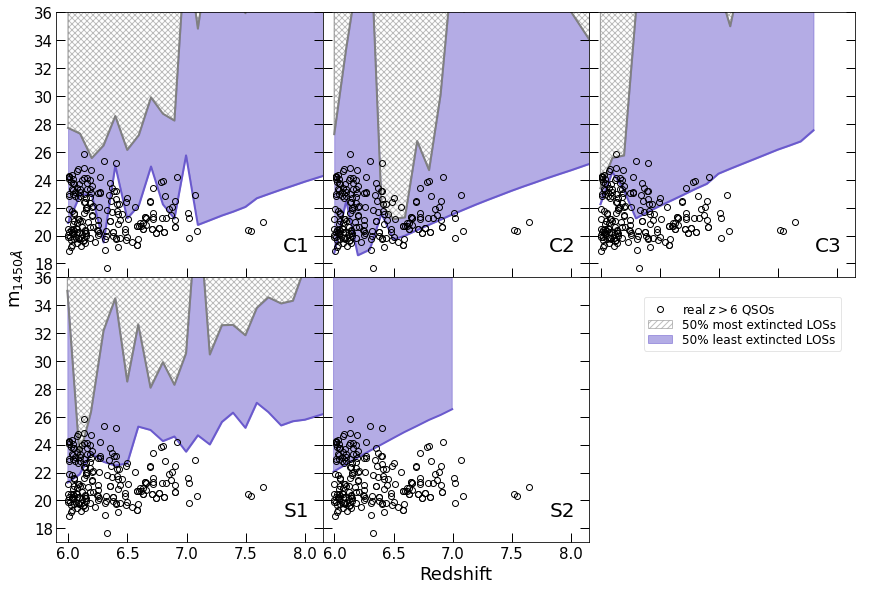}
    \caption{ Apparent magnitude at rest-frame $\lambda = 1450\,\angstrom$ as a function of redshift for bright AGN in the \cone (C1, C2, C3) and \sphere (S1, S2) simulations. The purple regions encompass the 50\% least extincted LOSs, while the grey hatched regions represent the 50\% most extincted LOSs. The grey circles are $z>6$ QSOs collected from \citet{Banados16, Banados18a}, \citet{Chehade18}, \citet{Matsuoka18a,Matsuoka18b, Matsuoka19a,Matsuoka19b}, \citet{Mazzucchelli17b}, \citet{Reed17}, \citet{Tang17}, \citet{Wang17, Wang18a,Wang18b, Wang19, Wang21b}, and \citet{Yang19,Yang20}.}
    \label{fig:m1450_z}
\end{figure*}

\subsection{UV magnitudes}\label{UV}
In \S~\ref{Xray_obsc} we discussed how the available X-ray observations of $z>6$ QSOs are not sensitive to the column density values that we derived for bright AGN in the simulations. Instead, the rest-frame UV emission of high-redshift AGN is expected to be severely affected by dust extinction even for low values of $N_H$. In this section, we compare the expected rest-frame UV magnitudes of bright AGN in the simulations with the observed values of known $z>6$ QSOs.

We assumed that the intrinsic (i.e., unextincted) rest-frame UV spectra of the AGN-hosting galaxies in the simulations are dominated by the AGN (i.e., we do not include stellar emission) and are well represented by the \cite{VandenBerk01} composite spectrum of type 1 QSOs, rescaled to their bolometric luminosity via the bolometric correction of \cite{Venemans16} and \cite{Decarli18}
\begin{equation}
\mathrm{log}\left(\frac{L_{bol}}{\mathrm{erg\,s^{-1}}}\right)=4.553+0.911\times \mathrm{log}\left(\frac{\lambda L_{\lambda}(1450\angstrom)}{\mathrm{erg\,s^{-1}}}\right).
\end{equation}

%\cite{DiMascia21b} performed  extensive radiative transfer calculations on the \cone snapshot at $z=6.3$. They found that the attenuation curve that matches best observational results is based on a Small Magellanic Cloud (SMC) extinction curve, modified such that small dust grains are removed. It is beyond the scope of this paper to perform such expensive calculations for all of the AGN and redshifts considered in this work. Therefore, while in principle UV attenuation depends on the dust distribution, 
We assumed a simple uniform slab of dust located in front of each AGN and an SMC extinction curve, and computed the measured rest-frame UV flux as

\begin{equation}
    F_\lambda^\mathrm{obs}=F_\lambda^\mathrm{intr}e^{-\tau_\lambda},
\end{equation}
where $\tau_\lambda=k_\lambda\Sigma_m f_{dust}$, $k_\lambda$ is the extinction cross section at wavelength $\lambda$, $\Sigma_m$ is the mass column density of metals, which we computed in \S~\ref{NH}, and the fraction of metal mass locked into dust is assumed to be $f_{dust}=0.15$ as in \cite{DiMascia21b}. Finally, we computed the apparent magnitude at the wavelength corresponding to rest-frame 1450 \angstrom, that is $m_{1450}$.

For all considered AGN, the metal mass is computed from the column densities of metals derived in \S~\ref{NH} for 1000 LOSs at every simulation snapshot. Thus, we obtain a distribution of 1000 values of $m_{1450}$ at every redshift. In Fig.~\ref{fig:m1450_z} we show the magnitudes obtained for the 50\% least (purple regions) and most (grey hatched regions) extincted LOSs. % We also report, for comparison, the un-extincted magnitudes as dotted black lines. 
To allow for a comparison with observations, we add the magnitudes of a sample of $z>6$ QSOs collected from \cite{Banados16, Banados18a}, \cite{Chehade18}, \cite{Matsuoka18a,Matsuoka18b, Matsuoka19a,Matsuoka19b}, \cite{Mazzucchelli17b}, \cite{Reed17}, \cite{Tang17}, \cite{Wang17, Wang18a,Wang18b, Wang19, Wang21b}, \cite{Yang19,Yang20}, with typical magnitudes of $19\lesssim m_{1450}\lesssim 24$.

Among the considered simulations, \cone produces the UV brightest AGN, which are consistent with the magnitudes of known QSOs at $z\lesssim7$. As discussed in \S~\ref{NH_distro}, such redshift range corresponds to the period when the AGN strong kinetic feedback strongly affects the gas column density in the host galaxy,  strongly suggesting that known, optically selected $z>6$ AGN are indeed observed preferentially along directions where AGN feedback has cleared the LOS of most of the gas and dust. 
 This prediction is hard to be tested observationally. Not only estimating the outflow direction is a difficult task, but the incidence of outflow in high-redshift AGN itself is still a matter of debate (e.g., \citealt{Maiolino12, Cicone15, Bischetti19, Novak20, Izumi21, Meyer22}). Moreover, $z>6$ QSOs might have been detected along LOSs which have been previously cleared of most of the gas and dust by past outflows. 
 In this respect, a caveat arises from the numerical implementation of the ISM properties in the \cite{Barai18} simulations, which, as described in \S~\ref{Numerical_methods}, follow the prescription of \cite{Springel03}. This model does not capture the ISM porosity and therefore is not able to resolve clumpy structures on ~pc scales. Resolving such structures might decrease the effective opacity of the medium and possibly produce more unobscured lines of sight, even in the absence of AGN feedback.

In Fig.~\ref{fig:m1450_z}, only $<50\%$ of the LOSs of an individual AGN have extinction values small enough to reproduce the observed magnitudes.
We computed the probability that multiple AGN  appear as UV bright (i.e., $m_{1450}\lesssim24$) sources along the same LOS, and found that it is negligible. This result is consistent with observations, according to which, to date, no such a system of multiple UV-bright AGN has been discovered at high redshift.

The most luminous AGN in the \sphere run, S1, reaches magnitudes as bright as the observed values only at $z\approx6.5$, while it fails at reproducing the magnitudes of $z>6.5$ QSOs. This is due to the lower accretion rate, and thus lower intrinsic luminosity, of S1 than the accretion rates of bright AGN in \cone. The large column density of S2 results in dramatic extinction levels along all of the LOSs, such that only along a small fraction of the LOSs S2 has apparent magnitude consistent with those of observed high-redshift QSOs, despite its intrinsic luminosity being similar to that of S1 at $z<6.5$ (Fig.~\ref{fig:mdot}).

\begin{figure*}
    \centering
    \includegraphics[width=1\textwidth ]{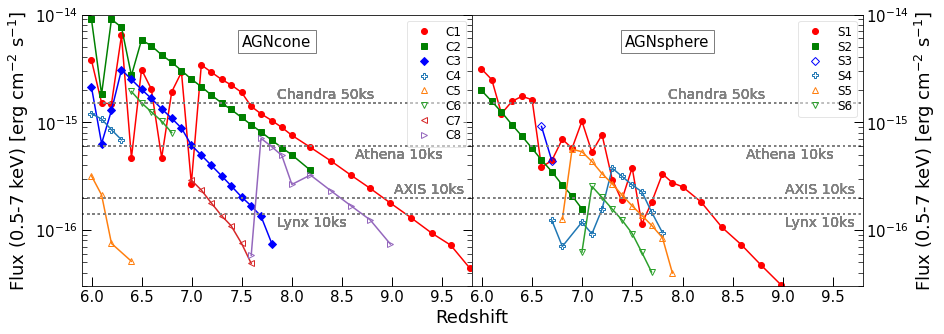}
    
    \caption{Expected X-ray flux in the 0.5--7 keV band as a function of redshift for the  \cone (\textit{left panel}), and \sphere (\textit{right panel}) runs. For each AGN at each redshift, we assumed the median $N_H$ computed for 1000 lines of sight. The horizontal dotted lines mark the flux limit computed for \chandra (50 ks observation), \athena (10 ks), \axis (10 ks), and \lynx (10 ks).  }
    \label{fig:Fx}
\end{figure*}

\section{Multiple high-redshift AGN on 1-10 arcsec scales}\label{environment}

 Typical separations between AGN in the \cite{Barai18} simulations are $\approx5-50$ kpc, corresponding to only a few arcseconds in projection. To date, no multiple AGN system has been discovered observationally at $z>6$ (e.g., \citealt{Greiner21}), with the highest redshift AGN pair being recently discovered at $z=5.7$ \citep{Yue21}. This result could be due to dust extinction preventing the detection of other possible accreting SMBHs close to high-redshift QSOs, as we found in our simulations (\S~\ref{UV}). Alternatively, QSOs observed at $z\gtrsim6$ intrinsically have no AGN satellite. The latter hypothesis implies that the simulations overpredict the number of  bright AGN, due to, e.g., the specific numerical setup and seeding prescription. In addition, as discussed in \S~\ref{Numerical_methods}, the simulations focus on an overdense region, which maximizes the probability of forming multiple SMBHs, and thus bright AGN, in a small volume. 

To investigate better the relation between the predicted and observed number of systems of multiple AGN at high redshift, 
in \S~\ref{mock_Xray} we produce mock X-ray observations with the \chandra X-ray observatory\footnote{\url{https://cxc.harvard.edu/}} based on the \cone and \sphere simulations. Then, in \S~\ref{multiple_AGN} we compute the probability of detecting multiple AGN on small angular separations, and compare the findings with observational results. Finally, in \S~\ref{future_facilities} we investigate the potential of future X-ray facilities in detecting possible multiple faint AGN over small scales around bright high-redshift QSOs.

\subsection{Mock X-ray observations}\label{mock_Xray}

As discussed in \S~\ref{Xray_obsc}, the column densities that we derived in \S~\ref{NH_distro} for simulated $z>6$ AGN have a negligible effect on the X-ray emission at the observed-frame energies probed by X-ray telescopes, allowing us to factor out the effect of varying $N_H$ along different LOSs. However, we have to take into account another effect related to the specific choice of the LOS: the emission of different AGN might be blended along some LOSs due to projection effects, and appear as a single X-ray source. This effect might be important as the projected angular separations of the AGN in the considered simulations are comparable with the angular resolution of \chandra (i.e., $\approx0.5\arcsec$), which is the existing X-ray observatory with the sharpest view.

We produce mock observations using the SOXS v. 3.0 software,\footnote{\url{https://hea-www.cfa.harvard.edu/soxs/}} using \chandra response matrices and ancillary files suitable for Cycle 20. SOXS accounts for three background components: a uniform Galactic component, a cosmic background due to point-like sources, and an instrumental component. For each simulation, we produce two sets of mock images, assuming an exposure time of 30 ks or 50 ks, which are typical lengths of real \chandra observations of $z>6$ QSOs \citep[e.g.,][]{Vito19a,Wang21a}.
For each set, we considered 100 random LOSs, along which all AGN have been projected on the sky plane according to their tri-dimensional positions in the simulations. This allows us to statistically take into account 1) the possible blending of multiple sources due to projection effects, and 2) the Poisson fluctuations of the number of detected X-ray photons at a given intrinsic flux. 

We convert the bolometric luminosities of AGN in the simulations into X-ray luminosities in the rest-frame $2-10$ keV energy band using the \cite{Duras20} relation.
Then, we compute the fluxes in the 0.5-7 keV band (i.e., one of the standard energy bands used to analyse  \chandra observations) for every AGN, and use them as input
values to simulate the images. We adopt intrinsic powerlaw emission with photon index $\Gamma=2$. This is a typical value for AGN up to $z\approx6.5$ \citep[e.g.][]{Nanni17,Vito19b}, although \cite{Vito19b} and \cite{Wang21a} find hints for a steepening at higher redshifts. We also include absorption due to the measured value of column density along the considered LOS, although, as discussed above, the produced obscuration is negligible for our high-redshift objects, and a Galactic absorption component with $N_H=5\times10^{20}\nhunit$. These computations have been performed with XSPEC v.12.11 (\citealt{Arnaud96}; model $phabs\times zvphabs\times powerlaw$)\footnote{\url{https://heasarc.gsfc.nasa.gov/xanadu/xspec/}}. Fig.~\ref{fig:Fx} presents the expected X-ray flux of every AGN in the simulations as a function of redshift. %An example of a \chandra simulated image is reported in the first column of Fig.~\ref{fig:sim_image}.

\subsection{X-ray detection of multiple AGN}\label{multiple_AGN}
We ran a blind source detection procedure on the \chandra mock observations in the 0.5-7 keV band using the \textit{wavdetect} tool in CIAO v.4.12\footnote{\url{https://cxc.harvard.edu/ciao4.12/}} \citep{Fruscione06}, with a significance threshold of $10^{-5}$, over an area corresponding to $<30$ kpc from the central QSO, to be consistent with the volume considered throughout this work (see \S~\ref{Method}). We repeated this procedure for all snapshots in the $z=6-7$ range, which includes most of the $z>6$ QSOs observed with \chandra, thus allowing for a fair comparison with real observations. 

\begin{figure}
    \centering
    \includegraphics[width=0.5\textwidth ]{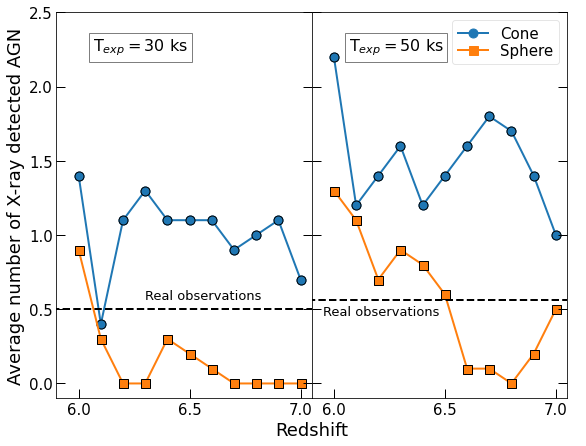}
    \caption{Average number of detected X-ray sources, averaged over 100 LOSs, detectable in the two simulations within $<30$ kpc from the central AGN with 30 ks (left) and 50 ks (right) \chandra observations. The black dashed line mark the average number of detected sources in real observations of $z>6$ QSOs. }
    \label{fig:Ndet}
\end{figure}

Fig.~\ref{fig:Ndet} presents the number of AGN detected in the mock \chandra observations with 30 ks and 50 ks exposures, averaged over the 100 LOSs, for each simulation. \cone predicts an average of $\approx1$ detectable AGN already with relatively short exposures (30 ks) and multiple detected X-ray sources using slightly longer observations (50 ks) over all of the considered redshift range. Instead,  according to the \sphere run, 30 ks (50 ks) \chandra observations of $z\gtrsim6.2$ ($z\gtrsim6.5$) should typically return no detected source, but the probability to detect one or more AGN increases quickly approaching $z=6$.

\begin{table*}
	\caption{Comparison sample of \chandra observations of $z=6-7$ QSOs (see \S~\ref{multiple_AGN}).}
		\begin{tabular}{cccccccccc} 
			\hline
			\multicolumn{1}{c}{{ ID }} &
			\multicolumn{1}{c}{{ z}} &			
			\multicolumn{1}{c}{{ Ref}} &
			\multicolumn{1}{c}{{ ObsID}} &			
			\multicolumn{1}{c}{{ $t_{exp}$ [ks] }} &
		\multicolumn{1}{c}{{ $N_{det}$}} 	\\ 
			\multicolumn{1}{c}{{ (1) }} &
			\multicolumn{1}{c}{{ (2)}} &			
			\multicolumn{1}{c}{{ (3)}} &
			\multicolumn{1}{c}{{ (4)}} &			
			\multicolumn{1}{c}{{ (5)}} &
		\multicolumn{1}{c}{{ (6)}} 	\\ 		
			%		(1) & (2) & (3) & (4) & (5) & (6) & (7) & (8) & (9) & (10)\\
			\hline
			\multicolumn{6}{c}{{ 20-40 ks sample}} \\
J002429.77+391319.0 & 6.621 & W21 & 20416 & 20 & 0 \\
J005006.67+344521.6 & 6.253 & V19 & 20393 & 34 & 1 \\
J022601.87+030259.4 & 6.541 & V19 & 20390 & 26 & 1 \\ 
J084229.43+121850.5	& 6.076 & V19 & 20392 & 29 & 0 \\
J104819.09-010940.2 & 6.676 & W21 & 20415 & 35 & 0 \\
J150941.78-174926.8 & 6.122 & V19 & 20391 & 27 & 1 \\
J152637.84-205000.7$^*$ & 6.586 & C20 & 22165 & 33 & 0\\
J163033.90+401209.7 & 6.065 & V19 & 5618  & 27 & 1 \\
			\multicolumn{6}{c}{{ 40-80 ks sample}} \\
J010953.13-304726.3 & 6.791 & V19 & 20398,22214 & 66 & 0\\
J030516.92-315055.9 & 6.614 & V19 & 20394 & 50 & 0 \\
J103027.11+052455.1$^*$ & 6.308 & N17 & 19926 & 50 & 1 \\
J111033.98-132945.6$^*$ & 6.515 & V19 & 20397 & 54 & 0\\
J114816.65+525150.4 & 6.419 & G17 & 17127 & 78 & 1 \\
J164121.73+375520.2 & 6.047 & V19 & 20396,21961 & 54 & 1 \\
J203210.0-211402.3$^*$ &6.24& C19 & 20470 & 45 & 1 \\
J223255.14+293032.3 & 6.666 & V19 & 20395 & 54 & 1 \\
J234833.34-305410.0 & 6.902 & W21 & 20414 & 42 & 0 \\
			\hline
		\end{tabular} \\\label{tab:highz_QSOs}
(1) ID of targeted QSO; (2) redshift of targeted QSO; (3) reference for published X-ray data. C19: \cite{Connor19}.  C20: \cite{Connor20}. G17: \cite{Gallerani17}. N17: \cite{Nanni17}. V19: \cite{Vito19b}. W21: \cite{Wang21a}. (4) \chandra observation ID considered in this work; (5) Exposure time; (6) number of detected X-ray sources according to the procedure described in \S~\ref{multiple_AGN}. $^*$ These QSOs have been observed with multiple ObsIDs, resulting in longer total exposure times than those reported here. We only consider the reported ObsIDs to allow for a fair comparison with our 30 ks and 50 ks mock observations. 
	\end{table*}

In order to compare these results with real data, we collected all of the available \chandra observations of $z=6-7$ QSOs with exposure times of 20-40 ks and 40-80 ks (Tab.~\ref{tab:highz_QSOs}). The median exposure time of the 20-40 ks (40-80 ks) observations is 38 ks (54 ks) and the median redshift of the targeted QSOs is $z=6.4$ ($z=6.5$). These values are well matched to our sets of 30 ks and 50 ks mock images, respectively. We repeated the detection procedure described above on the real \chandra observations, considering only an area of $R<30$ kpc from the targeted QSO, to allow for a fair comparison with the mock image results. We stress that the blind detection procedure prevents any bias related to rest-frame UV pre-selection of possible X-ray sources. 

The last column of Tab.~\ref{tab:highz_QSOs} reports the number of detected sources in the real observations,\footnote{We note that for almost all of the QSOs considered here, the results of the blind detection procedure agree with what reported in the literature, but for J084229.43+121850.5. \cite{Vito19b} reported a detection of X-ray emission from this QSO, while here we report it as undetected. This apparent discrepancy is due to the different detection procedure (i.e., blind detection vs.  rest-frame UV pre-selection of the target position) and  significance threshold.} which are almost equally split between no detected source and one detected source (i.e., the targeted QSO): the average numbers of detected X-ray sources in one observation are 0.50 and 0.56 for the 20-30 ks and 40-80 ks samples, respectively. Similar values are obtained by splitting each sample according to its median redshift.  Comparing these results with the expected numbers of detected sources in simulations (Fig.~\ref{fig:Ndet}), we find that \cone overestimates the number of detectable AGN at all redshifts, assuming both 30 ks and 50 ks exposure times. Instead, \sphere underestimates such number assuming 30 ks observations, while shows a strong dependence on redshift for longer exposures: at $z>6.5$ and $z<6.5$ it underestimates and overestimates, respectively, the average number of detected X-ray sources.

\begin{figure*}
    \centering
    \includegraphics[width=1\textwidth ]{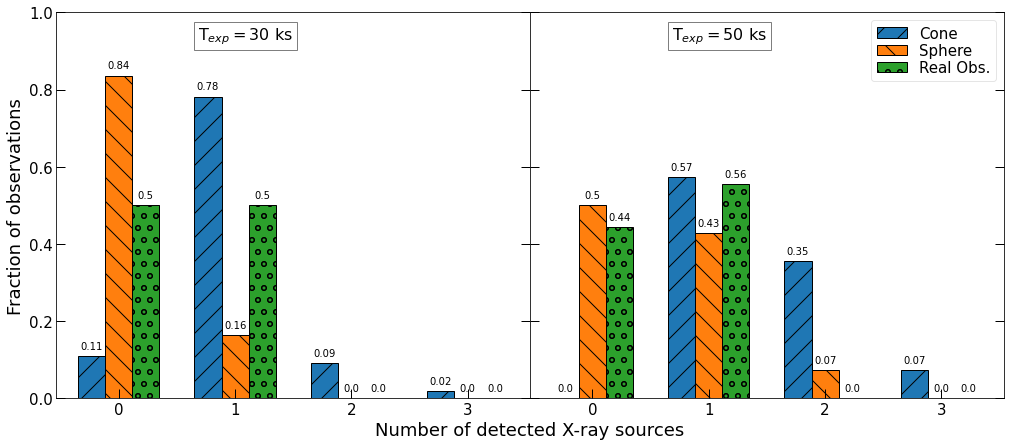}
    \caption{Normalized histograms of the number of detected X-ray sources in the mock and real \chandra observations of $z=6-7$ AGN, for $t_{exp}=$ 30 ks (left) and 50 ks (right).}
    \label{fig:Ndet_hist}
\end{figure*}

Due to the small sample sizes of real QSO observations and the narrow range covered by the number of detectable X-ray sources, it is difficult to provide a quantitatively robust comparison with the predictions from simulations. Nonetheless, we attempt to do it by comparing the normalized histograms of detected sources in the mock and real observations over the entire $z=6-7$ range  (Fig.~\ref{fig:Ndet_hist}). This is justified by the relatively flat redshift distribution of the QSOs targeted by real observations (Tab.~\ref{tab:highz_QSOs}). For each set of mock images, we computed the two-sample Anderson-Darling test.\footnote{We used the \textit{anderson\_ksamp} method of the SciPy package \citep{Scipy20}.} The null hypothesis is that the mock and real observations are drawn from the same parent population, for what the number of detected X-ray sources is concern.
We found that the null hypothesis can be rejected with high significance (i.e., Anderson-Darling test sigificance level $\lesssim0.001$) for almost all combinations of simulations and exposure times: Fig.~\ref{fig:Ndet_hist} confirms that \cone and \sphere  overestimate and underestimate, respectively, the number of detectable X-ray sources. Mock simulations of \sphere with $t_{exp}=50$ ks is the only set for which the null hypothesis cannot be rejected, although this simulation is not consistent with real observations for $t_{exp}=30$ ks.
It is worth noting that few $z>6$ QSOs have been pointed with long \chandra exposures (100--500 ks; e.g. \citealt{Nanni18}, \citealt{Connor20}, \citealt{Vito21}). Some of these observations were performed to check the presence of faint and possibly obscured AGN around $z>6$ QSOs, for which companion galaxies have been detected with ALMA and HST. However, to date, no solid detection of such satellite AGN has been obtained (\citealt{Vito19a,Vito21, Connor19,Connor20}).

\subsection{Predictions for future X-ray facilities}\label{future_facilities}
The high sensitivities of future X-ray facilities will allow us to push the search for AGN satellites of luminous optically selected QSOs at $z>6$ down to intrinsic luminosities significantly lower than those probed with \chandra. In Fig.~\ref{fig:Fx} we report as dotted grey lines the approximate expected sensitivity limits of future missions such as \athena/WFI \citep{Nandra13}, \axis \citep{Mushotzky19, Marchesi20}, and \lynx/HDXI \citep{Gaskin19}, each one computed assuming 10 ks exposure time, and compare them with the sensitivity of a 50 ks \chandra observation. We computed these values by simulating X-ray observations of an X-ray source, assuming a simple power-law spectrum with photon index $\Gamma=2$ and varying flux. In particular, for each instrument, we loaded response matrices and background files\footnote{We use real response matrices and background files for \chandra, and the preliminary files included in SOXS for \lynx, \axis, and \athena.} in XSPEC, and computed the expected source and background count rates in a region including $\approx90\%$ of the expected point spread function (PSF); i.e., $R=1\arcsec$ for \chandra, \axis, and \lynx, and $R=5\arcsec$ for \athena. Then, we computed the flux that returns a binomial no-source detection probability \citep[i.e., $P_B$;][]{Weisskopf07} such that $(1-P_B)=0.997$, corresponding to $3\sigma$ in the Gaussian approximation.

\begin{figure*}
    \centering
    \includegraphics[width=1\textwidth ]{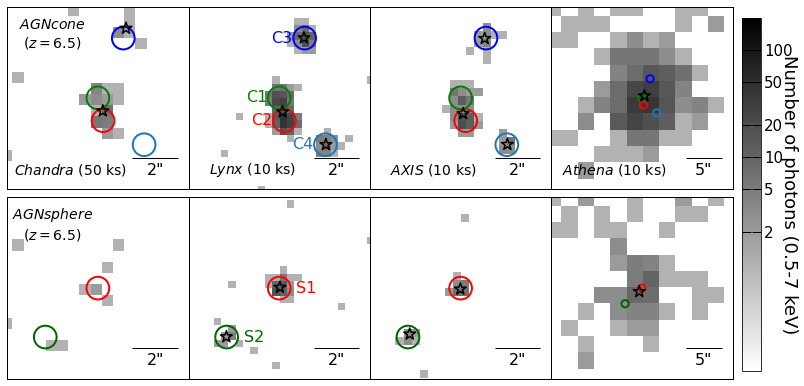}
    \caption{Simulated X-ray observations in the 0.5--7 keV band  of the most-massive AGN at $z=6.5$ and the surrounding satellite AGN in the \cone (upper row) and \sphere (lower row) simulations. From the leftmost to the rightmost columns, we simulated observations with \chandra/ACIS-S (50 ks), \lynx/HDXI (10 ks), \axis (10 ks), and \athena/WFI (10 ks). For presentation purpose, the angular scale of the \athena image is different from the other cases, due to the larger PSF. The circles mark the location of the simulated AGN for a representative line of sight, and are color coded as in Fig.~\ref{fig:masses}. The black stars mark the position of X-ray detected sources obtained with a blind detection procedure.}
    \label{fig:sim_image}
\end{figure*}

Fig.~\ref{fig:Fx} shows that all of the considered next-generation X-ray mission will provide us with a huge improvement in the capability of detecting faint AGN at $z>6$, including  satellite AGN around bright QSOs at $z>6$, in a fraction of the time of a typical \chandra observation. Fig.~\ref{fig:sim_image} presents simulated X-ray observations with \chandra (50 ks), \lynx (10 ks), \axis (10 ks), and \athena (10 ks) of a representative snapshot (i.e., $z=6.5$) and LOS of the two simulation runs. The satellite AGN will appear as multiple X-ray sources on a few arcsec scales. This implies that, in addition to high sensitivity, excellent angular resolution, such as that provided by \axis and \lynx, is required to detect them individually. To probe this issue, we performed a blind detection run with \textit{wavdetect} on these images, and compared the detected sources (black stars in Fig.~\ref{fig:sim_image}) with the input AGN (colored circles): the identification of close objects like C1 and C2 is difficult even with missions with $\approx0.5$ arcsec angular resolution. The problem is clearly more evident with \athena, due to its PSF of a few arcsec.

\section{Discussion}\label{discussion}
As discussed in \S~\ref{Numerical_methods},  the outflow directions in the considered simulations are assumed not to be physically related to the host-galaxy properties and to be time-independent. 
In particular, the \cone simulation does not assume the outflow to be perpendicular to the plane of the host galaxy, as suggested by several observations of kpc-scale outflows or radio jets in the local universe \citep[e.g.,][]{Garcia-Burillo14, Cresci15, Morganti15, Venturi21}, where the outflow geometry can be studied in details, and by some numerical simulations \citep[e.g.,][]{Hopkins12}.

Several physical mechanisms can concur in the acceleration of winds at sub-pc scales that eventually produce large-scale outflows, including magneto-hydrodynamic effect (e.g., \citealt{Sadowki13}), thermal driving (e.g., \citealt{Proga07}), radiation pressure acceleration, either applied on dust (e.g., \citealt{Ishibashi15}) or mediated by UV transitions \citep[e.g.][]{Proga04,mizumoto2021}, which might produce outflows with different geometries. Moreover, the outflow geometry might be affected by interactions with the surrounding environment as the outflow expands \citep[e.g.][]{Nelson19, talbot2021}, and might change with time. Cosmological simulations cannot describe in detail such a complex, and largely unknown, physics and evolution of outflows with relatively simple numerical recipes. 

The goal of this paper is to investigate the effect of two particular large-scale outflow geometries (i.e., a spherical outflow and a bi-conical outflow parametrized as described in \S~\ref{Numerical_methods}) on the observable properties of high-redshift AGN, regardless of the sub-grid physical mechanisms responsible for their acceleration. Extensive numerical simulations with identical initial conditions and physics except for the outflow parameters would be required to check whether and how the results are sensitive to different choices of the outflow parameters.

Kinetic feedback produced during the phases of fast accretion of SMBHs in the \cite{Barai18} simulations  has a significant impact on  the  surrounding  material and is required to match the predicted observable properties of bright AGN with observational results. One of the strongest piece of evidence is represented by the study of the gas extent in the AGN host galaxies (Fig.~\ref{fig:NH_radius}): the gas reservoirs in the \noagn case (i.e., in absence of AGN feedback) are always more compact than those derived from ALMA observations of $z>6$ QSOs (see also, e.g., \citealt{vanderVlugt19}). The effect of AGN feedback pushes the gas in the host galaxies to larger distances (i.e., up to a few kpc) from the centres, in agreement with observations (e.g., \citealt{Cicone15,Bischetti19, Venemans20, Izumi21}). Although other mechanisms related to AGN feedback may produce such an observable, by, for instance, preventing gas infall from large scales (e.g., \citealt{Trussler20}) or causing fluctuations in the gravitational potential, which may lead to a radial migration of the material (e.g., \citealt{vanderVlugt19}), \cite{Barai18} found that the mechanical removal of gas from the inner region of the host galaxies is the main process that affects their gas content in their simulations. 
We underline that also  some $5<z<7$ star-forming ($1-70\,\mathrm{M_\odot\, yr^{-1}}$) galaxies have been found to show both an extended [C II] halo \citep[e.g.,][]{Fujimoto20} and broad wings in the [C II] emission-line profile \citep[e.g.,][]{Gallerani18, Ginolfi20}, suggestive of outflows possibly powered by a yet undetected accreting MBH \citep[e.g.,][]{Orofino21}.

At $z<7$ the feedback produces a general decrease of the $N_H$  (Fig.~\ref{fig:NH_z_all}), allowing for the appearance of unobscured  (i.e., $N_H<10^{22}\,\nhunit$) LOSs (Fig.~\ref{fig:fLOS_all}). 
Such directions are most probably those along which known $z>6$ QSOs are preferentially observed, as the rest-frame UV selection of these objects requires low dust extinction. In fact, at $z\lesssim6.5$, when the feedback effect is the strongest, bright AGN in the \cone simulation are able to reach the UV magnitudes observed for known $z>6$ QSOs (Fig.~\ref{fig:m1450_z}). 
However, such LOSs represent only a fraction of the total LOSs of an AGN (see also, e.g., \citealt{Ni20, Trebitsch19, Lupi22}): more than half of the LOSs would appear too faint to be selected as high-redshift objects in current optical/near-IR surveys, suggesting that a large fraction of the high-redshift, intrinsically luminous QSO population is observationally missed due to strong UV extinction produced by the ISM only. The presence of a dusty torus on pc scales, which is not included in the simulations we have analysed, would further increase such a fraction.

The outflow geometry likely plays an important role: in the case of a conic outflow, SMBH accretion proceeds at maximum efficiency through equatorial infalling of gas until $\dot{M}\approx10-30\,\mathrm{M_\odot\,yr^{-1}}$ (Fig.~\ref{fig:mdot}), producing BHs with masses of $>10^9\,\mathrm{M_\odot}$ at $z=6-7$ (Fig.~\ref{fig:masses}). At these accretion rates, the feedback regulates further accretion and reduces the typical obscuring column density, in particular along the cone direction (Fig.~\ref{fig:Mollweide}).  In the case of outflows launched along random directions, the feedback can affect the growth of the SMBH and the $N_H$ distribution even at lower accretion rates, resulting in $<10^9\,\mathrm{M_\odot}$ BHs at $z=6$, provided that the gas in the host galaxy is not too dense, as in the case of S2. Thus, the ISM properties (i.e., $N_H$ and radial size of the gas) of the brightest AGN in the \sphere run is in agreement with observations. However, hindering the formation of $>10^{9}\,\mathrm{M_\odot}$ BHs, the spherical geometry of the feedback in \sphere prevents AGN from reaching intrinsic luminosities comparable to known $z>6$ QSOs at most redshifts (Fig.~\ref{fig:m1450_z}).

Interestingly, even the most luminous AGN in \cone cannot explain the detection of UV-bright QSOs at $z\approx7.5$ (Fig.~\ref{fig:m1450_z}), due to the combination of the relatively small BH masses, and hence low accretion rates, which, by construction, are capped at the Eddington rate, and typically high $N_H$ at that early cosmic time in this simulation. The existence of bright QSOs at $z\approx7.5$ \citep[e.g.,][]{Banados18a,Wang21a} requires different physical conditions for the SMBH formation and mass growth from those adopted in the considered simulations.\footnote{ As mentioned in \S~\ref{selection}, cosmic variance may affect our conclusions, as the simulations focus on a single cosmic region at high redshift.}  Future numerical simulations may explore such conditions as viable ways to reconcile the expected and observed properties of $z>7$ AGN. Non-mutually exclusive possibilities are: 

\noindent (a) different BH seeding mechanisms, that is, bright and massive QSOs discovered at $z\approx7.5$ may be grown from more massive BH seeds or have been seeded at earlier redshift than the SMBHs in the simulations.

\noindent (b) Sustained periods of super-Eddington accretion at $z>7.5$, whereas in the simulations the SMBH accretion rate is capped at the Eddington limit.

\noindent (c) Mass accretion characterized by a lower radiative efficiency than the value used in the simulations (i.e., $\epsilon_r=0.1$). In this case, the mass that is not converted into radiation contributes to the growth of SMBH, which can reach higher masses than those found in simulations at a given time. For instance, \cite{Davies19} report observational evidence for possible low radiation efficiency ($\epsilon_r\approx0.001$) in high-redshift QSOs.

 \noindent (d) High-redshift AGN typically reside in regions which are even more overdense than that investigated in the \cite{Barai18} simulations, thus favouring the formation of SMBHs at earlier epochs. However, this possibility would arguably make the discrepancy between the observed and expected number of multiple X-ray detected AGN on small scales even worse. In addition, observational studies return contradictory results on the typical large-scale environment of high-redshift AGN \citep[e.g., ][]{Ota18,Mazzucchelli19, Mignoli20,Overzier21}.

The analysis that we have performed demonstrates that the comparison between several observable properties of AGN predicted by the \cite{Barai18} simulations and the observational results, including both the properties of the individual galaxies and the environment,
can help us to validate the recipes and assumptions adopted in numerical simulations. In particular, we found that AGN in the considered simulations match the gas radial distributions and apparent UV magnitudes of high-redshift QSOs. In addition, the same set of simulations has been demonstrated to reproduce well a number of physical properties of $z>6$ QSOs, such as dust properties \citep{DiMascia21b}, multi-wavelength spectral energy distribution \citep{DiMascia21a}, and the number of UV-detected and [C II]-detected satellite galaxies (Zana et al. accepted).

However, we also found that the predicted number of X-ray detectable satellite AGN located over small scales around luminous high-redshift QSOs both in the \cone and \sphere simulations does not agree with the observational results. 
This observable is relatively easy to estimate from simulations as it depends primarily on the BH accretion rate only, once a suitable conversion to X-ray luminosity is assumed. Moreover, gas and dust absorption does not affect significantly the observed X-ray emission from high-redshift AGN, as opposed to UV emission, up to high column densities (log$\frac{N_\mathrm{H}}{\mathrm{cm^{-2}}}\approx23.5-24.0$; see \S~\ref{Xray_obsc} and \S~\ref{UV}).
 
The mismatch between the number of multiple X-ray detected AGN on small scales between simulations and observations  may be related to numerical issues and physical prescriptions. In particular, the simplistic BH seeding recipe implemented in the considered simulations (i.e., a $10^5\,\mathrm{M_\odot}$ BH is placed in the centre of a galaxy when this reaches a given mass threshold) naturally leads to the formation of a large number of SMBHs, that would appear as bright AGN at later cosmic times. Similar seeding recipies have been commonly adopted by most cosmological simulations (e.g.,  \citealt{Costa14}, \citealt{DiMatteo17}, \citealt{Barai18}, \citealt{Smidt18}, \citealt{Lupi19}, \citealt{Valentini21}), and typically mimic the ``heavy seed" formation channel for SMBHs \citep[e.g.,][]{Lodato06, Ferrara14}. However, theoretical models of ``heavy seed" formation require stringent physical conditions on, e.g., metallicity, physical state of the gas, ad radiation fields \citep[e.g.,][]{Ferrara14}. Accounting for such conditions in cosmological simulations is particularly difficult, but would reduce the number of formed SMBHs, and thus the discrepancy with observational results.
 
 Another possibility is that observed QSOs at high redshift do not reside in regions as dense as those probed in the analysed simulations (but see, e.g., Zana et al. accepted).  In this case, the formation of multiple SMBHs is expected to be hindered, helping us reconcile the expected number of X-ray sources with observational results. In addition, we would also expect to form less massive BHs, with direct consequences on the observational expectations discussed in this paper, as the BH mass is tightly linked with the maximum accretion rate, and thus AGN luminosity and feedback strength. Qualitatively, we would expect to derive fainter rest-frame UV and X-ray fluxes, weaker feedback, and, as a consequence (see Fig.~\ref{fig:NH_radius}), more compact gas reservoirs (i.e., similar to the \noagn case) than the values discussed in \S~\ref{radial_distro}, \S~\ref{UV}, and \S~\ref{environment}. 
Future X-ray facilities will provide us with the required sensitivity and angular resolution to investigate the presence of multiple faint AGN around bright high-redshift QSOs down to unprecedented flux limits (see \S~\ref{future_facilities}).

\section{Summary and conclusions} \label{conclusions}
We studied the observable properties of $z=6-10$ bright %(i.e., $L_{bol}>10^{46}\,\mathrm{erg\,s^{-1}}$ at $z=6$)
AGN in a suite of zoom-in cosmological simulations by \cite{Barai18} characterized by the inclusion of AGN kinetic feedback with either bi-conical (namely, \cone) and spherical (\sphere) outflow geometry. We focused our investigation on the gas column density and size in the host galaxies, the AGN rest-frame UV magnitude and X-ray fluxes, and the detectability of systems of multiple AGN over a few kpc scale in the X-ray band. We compared these quantities with a control simulation in which SMBHs are not seeded (i.e., \noagn), and observational results of $z>6$ AGN. We summarize our findings as follows.

\begin{itemize}
        \item \cone produces three bright AGN that grow up to $5\times10^8 < M_{\mathrm{BH}}<5\times10^9\,\mathrm{M_\odot}$ at $z=6$. These objects are characterized by a steady increase of their accretion rate up to $\approx10-30\,\mathrm{M_\odot\,yr^{-1}}$. Once such high values are reached (at $z\approx6.5-7$), the strong AGN feedback prevents further increase of the accretion rate. This behaviour is linked with the bi-conical geometry of the outflow, that allows steady infalling of material along the equatorial directions, at least until the feedback grows strong enough to affect most of the gas in the galaxy halo.
        In \sphere, the spherical geometry of the outflow affects gas accretion already at low and moderate SMBH growth rate. For this reason, the two bright AGN produced in \sphere reach lower values of BH masses (i.e., $2\times10^8 < M_{\mathrm{BH}}<5\times10^8\,\mathrm{M_\odot}$) and accretion rates ($\dot{M}<10\,\mathrm{M_\odot\,yr^{-1}}$) than objects in \cone.
        
        \item AGN host galaxies in \sphere have gas column densities of $N_H\approx10^{23}\,\mathrm{cm^{-2}}$ from their formation up to $z=6.5-7$, when $N_H$ presents a remarkable drop due to the strong AGN feedback. In fact, the $N_H$ in matched galaxies in \noagn continues to increase during the entire considered redshift range. The brightest AGN in \sphere presents a similar behaviour as those in \cone, although the $N_H$ is typically slightly lower. We interpret this difference again as due to the assumed spherical symmetry of the outflow. Instead, the second bright AGN in \sphere do not reach accretion rate sufficiently high to significantly affect the gas in the host galaxy.
        Our findings are consistent with the upper limits on $N_H$ recently reported for a set of $z>6$ AGN observed in the X-rays. 
        
        \item Kinetic feedback is required to match the gas extent reported for high-redshift QSOs (i.e., up to a few kpc). In fact, galaxies in \noagn present typical gas sizes of $<1$ kpc, while the extents of the gas reservoirs of AGN in \cone and \sphere increase up to the observed values of a few kpc at $z\lesssim7$. The exception is the second bright AGN in \sphere, due to its relatively low values of accretion rate.
        
        \item All AGN in the simulations would appear as obscured (i.e., $N_H>10^{22}\,\mathrm{cm^{-2}}$) along all lines of sight (LOSs) at $z>7$. These objects would be missed by currently employed UV-based selection methods, which are heavily affected by dust extiction, and would require observations in different bands (e.g., X-ray or infrared) to be unveiled. At later cosmic times, a fraction of LOSs (up to $\approx80\%$, depending on the specific AGN and redshift) have $N_H<10^{22}\,\mathrm{cm^{-2}}$. These are the preferential directions along which known, UV-selected $z>6$ QSOs are observed.
        
        \item Under simple, but reasonable, assumptions on the gas-to-dust mass scaling and dust distribution, we estimate the apparent UV magnitudes ($m_{1450}$) of the AGN in the simulations along different LOSs. We found that AGN in \cone have $m_{1450}$ consistent with those observed for real high-redshift QSOs (i.e., $m_{1450}<25$) along $\lesssim50\%$  of the LOSs at $z<7$. AGN in \sphere, instead, have fainter magnitudes, due to the lower instrinsic luminosities, and, for the second AGN, the high extinction levels along most of the LOSs. No AGN in the simulations can reproduce the observed UV magnitudes of the few $z\approx7.5$ QSOs known to date, whose formation and accretion history are likely not well captured by the prescriptions assumed in the simulations.
        
        \item The presence of multiple bright AGN over scales of a few kpc led us to investigate their detectability in X-ray observations with \chandra, and to compare the results with real observations of $z>6$ QSOs. We found that the \cone run significantly overpredicts the number of X-ray detected multiple AGN at high redshift. Instead, \sphere produces AGN with lower rate of X-ray detection than typical values derived in relatively shallow (i.e., $30$ ks) observations, while it is consistent with the results obtained with longer (i.e., $50$ ks) observations.
        
\end{itemize}
These results demonstrate that the AGN in the considered simulations have physical properties consistent with those of real QSOs for what the column density and extent of the gas in the host galaxies and the UV magnitudes are concerned. A bi-conical geometry for the outflow is favored over a spherical geometry, as it reproduces AGN with the high luminosities and SMBH masses observed for $z=6-7$ QSOs. However, both simulations cannot explain the recent discovery of luminous QSOs at $z\approx7.5$, which may have been formed at higher redshift than the assumed seeding time in our simulations, or may have undergone extensive periods of super-Eddington accretion.

Moreover, we showed that the number of multiple AGN detectable in X-ray band over few kpc scales is the observable property that the considered simulations struggled the most to reproduce. We propose that this issue can be due to the  simplistic BH seeding methods generally implemented in cosmological simulations, that do not account for the complex physics related with the formation and rapid growth of massive BHs in the early Universe. Future X-ray observatories will provide us with the sensitivity required to investigate the possible presence of multiple faint AGN satellites around luminous QSOs at high redshift.

\section*{acknowledgements}
 We thank the anonymous referee for their valuable comments.
This research has made use of data obtained from the Chandra Data Archive and the Chandra Source Catalog, and software provided by the Chandra X-ray Center (CXC) in the application packages CIAO and Sherpa.
This research made use of Astropy,\footnote{\url{http://www.astropy.org}} a community-developed core Python package for Astronomy \citep{Astropy13, Astropy18}. SG acknowledges support from the PRIN-MIUR 2017 grant (PI Fabrizio Fiore).

\section*{Data Availability Statement}
The data underlying this article will be shared on reasonable request to the corresponding author.

\bibliographystyle{mnras}
% \bibliography{file_bibliography/ref}
\bibliography{file_bibliography/biblio}
\bsp
\label{lastpage}

\end{document}